\documentclass[fleqn,usenatbib]{mnras}

\usepackage[T1]{fontenc}
\usepackage{graphicx}
\usepackage{xcolor}
\usepackage{color}
\usepackage{hyperref}
\usepackage{notes2bib}
\usepackage{array}
\usepackage{graphicx} 
\usepackage{multirow}
\usepackage{hhline,booktabs}
\usepackage{makecell}
\usepackage{amsmath,amssymb}
\usepackage{subcaption}
\usepackage{newtxtext,newtxmath}
\usepackage{longtable}
\usepackage{pdflscape}
\usepackage{float}
\usepackage{placeins}
\hypersetup{colorlinks=true,linkcolor=[rgb]{1.,0.2,0.2},citecolor=[rgb]{0.1,0.4,1.},filecolor=[rgb]{0.7,0.2,0.2},urlcolor=[rgb]{0.0,0.2,1.}}
\makeatletter
\@fleqnfalse
\@mathmargin\@centering
\makeatother

\def\INSPIRE{\mbox{{\tt INSPIRE}}}
\newcommand{\Reff}{$\mathit{R}_{\mathrm{e}\,}$}
\newcommand{\Mstar}{${M}_{\star}\,$}

\newcommand{\kms}{km s$^{-1}$}

\usepackage[normalem]{ulem}

\definecolor{amber}{rgb}{1.0, 0.49, 0.0}

\newcommand{\comm}[1]{}
\definecolor{darkgreen}{rgb}{0.09, 0.45, 0.27}

\defcitealias{Ferre-Mateu+17}{FM17}
\defcitealias{Spiniello23}{\INSPIRE\, DR3}
\defcitealias{Maksymowicz-Maciata24}{MM24}
\title[INSPIRE VII. The local environment]{ \centering INSPIRE: INvestigating Stellar Population In RElics - VII. \\ The local environment of ultra-compact massive galaxies}

\author[D. Scognamiglio et al.]{\noindent
Diana Scognamiglio$^{1}$\thanks{E-mail: dianas@jpl.nasa.gov},
Chiara Spiniello$^{2,3}$, 
Mario Radovich$^{4}$, Crescenzo Tortora$^{3}$, Nicola R.~Napolitano$^{5}$, \and Rui Li$^{6}$, Matteo Maturi$^{7}$,  
Michalina Maksymowicz-Maciata$^{2}$, 
Michele Cappellari$^{2}$, 
Magda Arnaboldi$^{8}$,\and
Davide Bevacqua$^{9,10}$,
Lodovico Coccato$^{8}$,
Giuseppe D'Ago$^{11,3}$, 
Hai-Cheng Feng$^{12}$, 
Anna Ferr\'e-Mateu$^{13,14}$, \and
Johanna Hartke$^{15,16}$,
Ignacio Martín-Navarro$^{13,14}$,
Claudia Pulsoni$^{17}$
\\ 
$^{1}$Jet Propulsion Laboratory, California Institute of Technology, 4800,  Oak Grove Drive - Pasadena, CA 91109, USA\\
$^{2}$Sub-Dep. of Astrophysics, Dep. of Physics, University of Oxford, Denys Wilkinson Building, Keble Road, Oxford OX1 3RH, United Kingdom\\
$^{3}$INAF -  Osservatorio Astronomico di Capodimonte, Via Moiariello  16, 80131, Naples, Italy\\
$^{4}$INAF - Osservatorio astronomico di Padova, Vicolo Osservatorio 5, I-35122 Padova, Italy\\
$^{5}$ Department of Physics “E. Pancini”
University of Naples Federico II
C.U. di Monte Sant’Angelo
Via Cintia ed. 6, 80126 Naples, Italy\\
$^{6}$School of Physics, Zhengzhou University, Zhengzhou, 450001, China\\
$^{7}$ Zentrum für Astronomie, Universität Heidelberg, Philosophenweg 12, D-69120 Heidelberg, Germany\\
$^{8}$European Southern Observatory,  Karl-Schwarzschild-Stra\ss{}e 2, 85748, Garching, Germany\\
$^{9}$INAF - Osservatorio Astronomico di Brera, via Brera 28, 20121 Milano, Italy\\
$^{10}$DiSAT, Universitá degli Studi dell’Insubria, via Valleggio 11, I-22100 Como, Italy\\
$^{11}$Institute of Astronomy, University of Cambridge, Madingley Road, Cambridge CB3 0HA, United Kingdom \\
$^{12}$Yunnan Observatories, Chinese Academy of Sciences, Kunming 650216, Yunnan, People's Republic of China\\
$^{13}$Instituto de Astrof\'isica de Canarias, V\'ia L\'actea s/n, E-38205 La Laguna, Tenerife, Spain\\
$^{14}$Departamento de Astrofisica, Universidad de La Laguna, E-38200, La Laguna, Tenerife, Spain\\
$^{15}$Finnish Centre for Astronomy with ESO (FINCA), FI-20014 University of Turku, Finland\\
$^{16}$Tuorla Observatory, Department of Physics and Astronomy, FI-20014 University of Turku, Finland\\
$^{17}$Max-Planck-Institut f\"{u}r  extraterrestrische Physik, Giessenbachstrasse, 85748 Garching, Germany}

\date{Accepted XXX. Received YYY; in original form ZZZ}

\pubyear{2024}

\begin{document}
\label{firstpage}
\pagerange{\pageref{firstpage}--\pageref{lastpage}}
\maketitle

\begin{abstract}
Relic galaxies, the oldest ultra-compact massive galaxies (UCMGs), contain almost exclusively ``pristine'' stars formed during an intense star formation (SF) burst at high redshift. As such, they allow us to study in detail the early mechanism of galaxy assembly in the Universe. Using the largest catalogue of spectroscopically confirmed UCMGs for which a \textit{degree of relicness} (DoR) had been estimated, the \INSPIRE\ catalogue, we investigate whether or not relics prefer dense environments.
The objective of this study is to determine if the DoR, which measures how extreme the SF history was, and the surrounding environment are correlated. In order to achieve this goal, we employ the AMICO galaxy cluster catalogue to compute the probability for a galaxy to be a member of a cluster, and measure the local density around each UCMG using machine learning-based photometric redshifts. We find that UCMGs can reside both in clusters and in the field, but objects with very low DoR ($<0.3$, i.e., a relatively extended SF history) prefer under-dense environments. We additionally report a correlation between the DoR and the distance from the cluster centre: more extreme relics, when located in clusters, tend to occupy the more central regions of them. We finally outline potential evolution scenarios for UCMGs at different DoR to reconcile their presence in both clusters and field environments. 
\end{abstract}




\section{Introduction}
In the $\Lambda$-Cold Dark Matter ($\Lambda$CDM) 
formation scenario, the formation of massive early-type galaxies (ETGs) is consistent with a two-stage formation model \citep{Oser+10, Naab+14}. 
In the first stage (at $z>2$), the central ``bulk'' of mass is formed via an intense and very fast starburst that quickly ends leaving an ultra-compact quiescent galaxy, known as red nugget \citep{Damjanov+09, Damjanov+15_compacts}. In a subsequent accretion phase, which is much more extended in time,  mergers and gas inflows cause a dramatic size growth but only minor mass change \citep{Daddi+05,Trujillo+07, vanDokkum08}.  Although the accreted material is preferentially assembled on the outskirts of a red nugget, it nevertheless contaminates, along the line-of-sight, the “in-situ”, pristine component that encodes the information about high-$z$ baryonic processes, thus affecting its spatial and orbital distributions. 

The stochastic nature of merging processes suggests that a non-negligible number of red nuggets at low-$z$ should exist, having slipped through cosmic time without interacting with other systems, thus not changing their stellar populations. These very old, red and ultra-compact nearby systems are called \textit{relic galaxies} \citep{Trujillo+09_superdense, Trujillo+14, Ferre-Mateu+17}, as they still bear the memory of the early conditions in which they formed. 
Therefore, relics offer unprecedented insights into the high-$z$ processes shaping galaxy formation and mass assembly with high precision, comparable to the study of nearby galaxies.  
Moreover, the number density of relics and its time evolution strongly depends on the physical processes shaping the size and mass evolution of galaxies, e.g. major and minor galaxy mergers and their relative importance, adiabatic expansion driven by stellar mass loss and/or strong feedback \citep{Quilis_Trujillo13, Furlong+15b, Wellons16, Flores-Freitas22, Moura2024}. Hence, finding and precisely counting relics in redshift bins is a very valuable way to constrain the physical scenarios driving the formation and size-evolution of massive ETGs.  

It is widely acknowledged that the properties of a galaxy are significantly influenced by its surrounding environment. The past merger history should manifest in the size of ETGs, with high-density environments favouring rapid growth through dry merging (\citealt{Nipoti_2009}; \citealt{vanDokkum+10}). However, studies have yielded mixed results regarding the correlation between ETG size and environment, both at intermediate redshifts and in the local Universe \citep{Kaviraj2011, Huertas-Company+13, Cappellari13, Hou16}. 
When restricting to UCMGs, without age and star formation histories (SFHs) distinction (i.e., relics and non-relics) the situation is unclear, with some works finding a higher number density in clusters than in the field \citep{Poggianti+13, Stringer+15_compacts}. However, this might be attributed to the fact that the majority of these studies have focused on massive objects which are expected to be more common in denser environments at any size. For instance, \citet{Tortora20} have performed a statistical analysis of the local environment of photometrically selected
ultra-compact (\Reff $<1.5$ kpc) and massive ($M_{\star}>8\times 10^{10}$ $M_{\odot}$) galaxies compared to normal-sized galaxies of similar stellar masses and colours. They have shown that the number density of UCMGs is higher in clusters only because the parent population they are derived from, i.e., red and massive ETGs, are more frequently found in these dense environments. This is also consistent to what is reported in \citet{Damjanov+15_env_compacts}. However, \citet{Tortora20} have also found that the fraction of UCMGs, calculated with respect
to the total parent population in the field is slightly higher compared to that in clusters (see right panel of fig.\,1 in \citealt{Tortora20}).  

Relic galaxies, i.e., the oldest UCMGs, containing almost exclusively stars formed during the first phase of the formation scenario, intuitively could be expected to be found in low-density environments with less hot gas in the intra-cluster medium (ICM). However, they have been observed also in clusters both in the local Universe 
(\citealt{Ferre-Mateu+17}, hereafter \citetalias{Ferre-Mateu+17})  and  up to $z\sim0.7$ \citep{Siudek2023}. Hydro-dynamical simulations have reached very similar conclusions, with relics being identified in clusters and field environments \citep{PeraltadeArriba+16, Flores-Freitas22, Kimmig23, Moura2024}. Moreover, both \citet{Flores-Freitas22}  and \citet{Moura2024}, analysing relics in the Illustris TNG50 simulation, have concluded that at $z=0$ they are closely connected to the environment in which their progenitors evolved. In particular, the progenitors of relics have been found to live in consistently higher density environments already at $z \ge 2$, while younger UCMGs residing in clusters were brought to them at a later cosmic time. However, \citet{Kimmig23}, analysing 36 quenched galaxies of stellar mass larger than $3\times10^{10}$ M$_{\odot}$ at $z=3.42$ from the Magneticum Pathfinder simulations, reached an opposite conclusion. They have found that these objects do not inhabit the densest nodes of the cosmic web, but rather sit in local under-densities. 

From an observation point of view, \citetalias{Ferre-Mateu+17} have hinted for a possible correlation between the environment and structural, kinematics and stellar population parameters. Among the three local massive relics analysed, one was found in the field, one in a small group, and one in a cluster. 
Interestingly, the structural, dynamical, and stellar population properties (size, mass, and SFH) seem to be more extreme for the relic in the centre of a large cluster (NGC~1277), intermediate for the relic living in the outskirts of a small group of galaxies (PGC 032873), and less extreme for the one in isolation (Mrk 1216). 
This result is however based on only three objects. Now, leveraging the large dataset built by \INSPIRE\  \citep{Spiniello20_Pilot},  we can extend this investigation, also pushing the redshift boundaries outside the local Universe.  
 
Understanding whether systematic differences exist between UCMGs with different SFHs and living in different environments is fundamental to shed light on their origin and evolution. This is the primary objective of this study, which is the seventh of the \INSPIRE\ series. In particular, we aim to determine the potential correlation between the local environment of UCMGs and the `\textit{degree of relicness}' (DoR), qualitatively introduced in \citetalias{Ferre-Mateu+17} and quantified in \citet{Spiniello23}, hereafter \citetalias{Spiniello23}.  We define the environment based on local galaxy density through two distinct methodologies. Firstly, a cluster search is conducted utilising the Adaptive Matched Identifier of Clustered Objects (AMICO; \citealt{Bellagamba18, Maturi19}). Secondly, the Galaxy morphoto-Z with neural Networks (GaZNets; \citealt{Li22}) is employed for precise determination of photometric redshifts and galaxy identification. Subsequent analysis explores the correlation between the environment and the DoR.

The paper is organised as follows. 
In section~\ref{sec:data}, we begin by presenting the data used, starting with a brief summary of the \INSPIRE\ Survey and its dataset,  
followed by a description of the two catalogues employed to identify clusters, the AMICO galaxy cluster sample and the GaZNets catalogue. In section~\ref{sec:analysis}, the main analysis to measure the density of the local environment for the UMCGs is outlined. 
Section~\ref{sec:results} presents our main findings, delving into the characterisation of the local environment and its correlation with the DoR.  We discuss the results in Section~\ref{sec:discussion}, also trying to relate them to possible formation scenarios. Finally, Section~\ref{sec:conclusion} provides a summary and conclusion for the paper.

\section{Data}
\label{sec:data}
Thanks to data collected as part of the ESO Large Observational program (ID: 1104.B-0370, October 2019-March 2023, PI: C. Spiniello), the \INSPIRE\ project \citep{Spiniello20_Pilot} has built the first catalogue of spectroscopically confirmed relics outside the local Universe ($0.1<z<0.5$), characterising their kinematics \citep{DAgo23}, stellar populations \citep{Spiniello+21,Spiniello23}, and low-mass end of the Initial Mass Function (IMF; \citealt{Martin-Navarro+23, Maksymowicz-Maciata24}, hereafter \citetalias{Maksymowicz-Maciata24}).

Here, we use the final \INSPIRE\ catalogue, presented in \citetalias{Spiniello23}. It comprises 52 UCMGs 
that were originally identified from the Kilo Degree Survey (KiDS;  \citealt{Kuijken11}) DR3 footprint \citep{deJong+17_KiDS_DR3} via a dedicated campaign \citep{Tortora+16_compacts_KiDS, Tortora+18_UCMGs, Scognamiglio20}.  
Among these, 38 have been confirmed as relics, as they have formed more than 75\% of their stellar mass during the first phase of the formation scenario (at $z>2$).  Moreover, as introduced earlier, for each of the 52 UCMGs a DoR has been computed. This is a dimensionless parameter, ranging from 0 to 1 and defined as 
\begin{equation}
    \mathrm{DoR}=\left [ f_{M_{\mathrm{tBB=3}}^{\star}} + \frac{0.5\mathrm{Gyr}}{t_{75}} + \frac{0.7\mathrm{Gyr} + \left ( t_{\mathrm{Uni}}-t_{\mathrm{fin}} \right )}{t_{\mathrm{Uni}}}\right ]\times \frac{1}{3},
\end{equation}
where $f_{M_{\mathrm{tBB=3}}^{\star}}$ is the fraction of stellar mass formed by $z=2$, $t_{75}$ is the cosmic time at which a galaxy has assembled 75\%  of its mass, $t_{\mathrm{fin}}$ is the final assembly time, when 100\% of the stellar mass is in place. Finally, $t_{\mathrm{Uni}}$ is the age of the Universe at the redshift of the object. 
Essentially, a higher DoR indicates an earlier and more rapid mass assembly, with the most extreme relics, that have formed the entire totality of their stellar masses at $z>2$, approaching a value of 1\footnote{Using the definition above, NGC~1277, the most extreme relics fully characterised in the local Universe and residing in a cluster, has a DoR $\sim0.95$.}. Conversely, UCMGs with a DoR of 0 have likely undergone a very prolonged star formation (SF), and have just stopped forming stars.

The \INSPIRE\ UCMGs span a wide range of DoR, from 0.06 to 0.83, although having very similar sizes, $0.5\le$ \Reff$ \le1.7$ kpc, stellar masses, $0.64\times10^{11}\le $ $M_{\star} \le 2.71\times10^{11}$$M_{\odot}$, and colours, $1.8\le (g-i)\le 2.3$. 
The DoR has allowed to split the 52 \INSPIRE\ UCMGs in three main families: extreme relics (DoR $>0.7$), these that have formed the totality of their stellar mass by $z=2$, relics ($0.34\le$ DoR $\le0.7$) which had formed at least 75\% of their stellar mass by $z=2$, and non-relics (DoR $< 0.34$) characterised by a more extended SFH.

From a stellar populations point of view, by definition, the DoR correlates with the integrated stellar age. A strong correlation is also found with stellar metallicities and a mild one with the [Mg/Fe]:  objects with a higher DoR have overall larger [M/H] and slightly larger [Mg/Fe] (see \citetalias{Spiniello23}).  
Moreover, it appears that the low-mass end of the IMF slope also correlates with the epoch of the SF (\citealt{Martin-Navarro+23}, \citetalias{Maksymowicz-Maciata24}). Finally, relics have systematically larger velocity dispersion values than non-relics of similar stellar mass, both normal-sized and ultra-compact (\citetalias{Spiniello23, Maksymowicz-Maciata24}). 

\subsection{The AMICO galaxy cluster catalogue}
\label{sec:AMICO}
The Adaptive Matched Identifier of Clustered Objects (AMICO; \citealt{Bellagamba18, Bellagamba19, Maturi19}) is an algorithm for the detection of galaxy clusters in photometric surveys, based on the Optimal Filtering technique. It allows to maximise the signal-to-noise ratio (SNR) of the clusters \citep{Maturi05} taking into account the luminosity, spatial distribution and photometric redshifts of galaxies. 
Briefly, AMICO searches for cluster candidates by convolving the 3D galaxy distribution with a redshift-dependent filter, which is defined as the ratio of a cluster signal, modelled with an analytical recipe, and a noise model derived directly from the data. Bayesian photo-$z$ \citep[BPZ;][]{Benitez00}, estimated from a template-fitting method, are used here. AMICO thus creates a 3D amplitude map where the candidate clusters are identified as peaks through an iterative approach designed to minimise the blending between nearby objects. The angular positions, redshift, signal amplitude, measuring the cluster galaxy abundance, and the SNR are retrieved for each cluster candidate. The mass of the cluster, based on weak lensing (WL) scaling relations, is derived too. Finally, the algorithm provides a probabilistic membership association of galaxies to clusters by exploiting the probability redshift distribution of each galaxy (provided by BPZ redshifts) and the model used for the cluster detection.
The cluster model is described by a luminosity function and a radial density profile \citep{Bellagamba18},  and observationally derived from the galaxy population of clusters detected through the Sunyaev--Zeldovich (SZ) effect \citep{Hennig17}.

The AMICO catalogue based on KiDS DR3 data \citep{deJong+17_KiDS_DR3} was  presented in \citet{Maturi19}: it covers an area of $414$ deg$^2$, and comprises 7988 candidate galaxy clusters at $0.1 < z < 0.8$. It  has been successfully used both to derive the population properties of galaxies in the identified clusters \citep{Radovich20,Puddu21}, as well as for WL \citep{Bellagamba19,Ingoglia22}, and cosmological analyses \citep{Giocoli21, Lesci22, Lesci22b, Romanello23}.

Here, we use the newest catalogue derived, applying the same algorithm on the KiDS DR4 \citep{Kuijken19_KIDSDR4}.
This catalogue, detailed in Maturi et al. (in preparation), spans a total area of 1006 deg$^2$, which, after masking, translates to an effective area of 840 deg$^2$. It includes 22\,614 candidate galaxy clusters within the photometric redshift range of $0.1 < z < 0.8$, detected down to a SNR $> 3.5$. The catalogue has an average purity of approximately 80\% across the entire redshift range. However, it is worth noting that purity strongly depends on the detection SNR. We refer the readers to Maturi et al. (in preparation) for a more quantitative description of the catalogue.

\subsection{The GaZNets catalogue}
\label{sec:gaznets}
The Galaxy morphoto-Z with neural Networks (GaZNets), introduced in \citet{Li22}, is a deep learning (DL) tool that combines both images and multi-band photometry measurements for the accurate determination of photometric redshifts. What makes this tool distinctive is its integration of conventional machine learning (ML) regression tools with DL techniques. GaZNets has been already successfully applied to a sample of galaxies from the KiDS DR4 \citep{Kuijken19_KIDSDR4}, and tested against other ML based catalogues and classification algorithms \citep{Khramtsov19}. 

In this work, we use as input the reference network developed by \citet{Li22}, GaZNet-1, which makes use of a combination of  KiDS $r$-band images and the KiDS+VIKING 9-bands catalogue ($ugriZYJHKs$, \citealt{Wright+18}). 
Of the 65.9 million sources in the original catalogue, roughly 40 millions have a measurement in all the bands, a high-precision photometric redshift ($z_{\rm phot}$), with uncertainty $\Delta z=0.038(1+z)$, 
and a corresponding probability of being classified as quasar, galaxy, or star. We use here the results obtained in Feng et al. (submitted), who classified $\sim27.3$ millions sources  with $r$-band magnitude $\le23$ from the KiDS DR5 database in quasars, galaxies, or stars using a multi-modal neural network. 
In particular, from the classification catalogue, we pre-select only $\sim 7.9$ million objects that have a very high probability to being a galaxy, $P_{\rm gal}\ge 0.9$, in the redshift range $0.1<z<0.5$. This threshold for the $P_{\rm gal}$ maximizes the completeness while minimizing the contamination.  Indeed, using $\sim20\,000$ galaxies with a spectroscopic match, \citet{Li22} have shown that, for both $P_{\rm gal}\ge 0.5$ and $P_{\rm gal}\ge 0.9$ probabilities, 99\% of the galaxies are classified correctly. Instead, increasing $P_{\rm gal}$ to 0.99,
only the 78\% of the galaxies will be classified as such. Moreover, considering $P_{\rm gal}\ge 0.5$, 4.6\% of the quasi-stellar objects (QSOs) are classified as galaxies. Indeed, using $P_{\rm gal}\ge 0.5$ we would retrieve $\sim8.5$ millions of galaxies, hence hinting at a larger contamination. We nevertheless caution the reader that because the spectroscopic sample is brighter, it is easier to distinguish between quasars, galaxies, and stars.

\section{Analysis}
\label{sec:analysis}
In this section, we conduct an in-depth analysis of the local environment for the \INSPIRE\ UCMGs from two perspectives.
Firstly, we investigate their potential association with galaxy cluster candidates by cross-referencing \INSPIRE\ catalogue data with the AMICO catalogue. We focus on evaluating the probability of UCMGs being members of these clusters and its possible correlation with the DoR. Secondly, we measure the local density  
around each \INSPIRE\ UCMG by conducting a cross-match with the GaZNets catalogue and counting galaxies with compatible redshifts. Through this analysis, we aim to determine whether a statistically significant over-density exists, thereby indicating a cluster environment, and its correlation with the DoR. 

\begin{figure}
    \centering  \includegraphics[width=1.08\columnwidth]{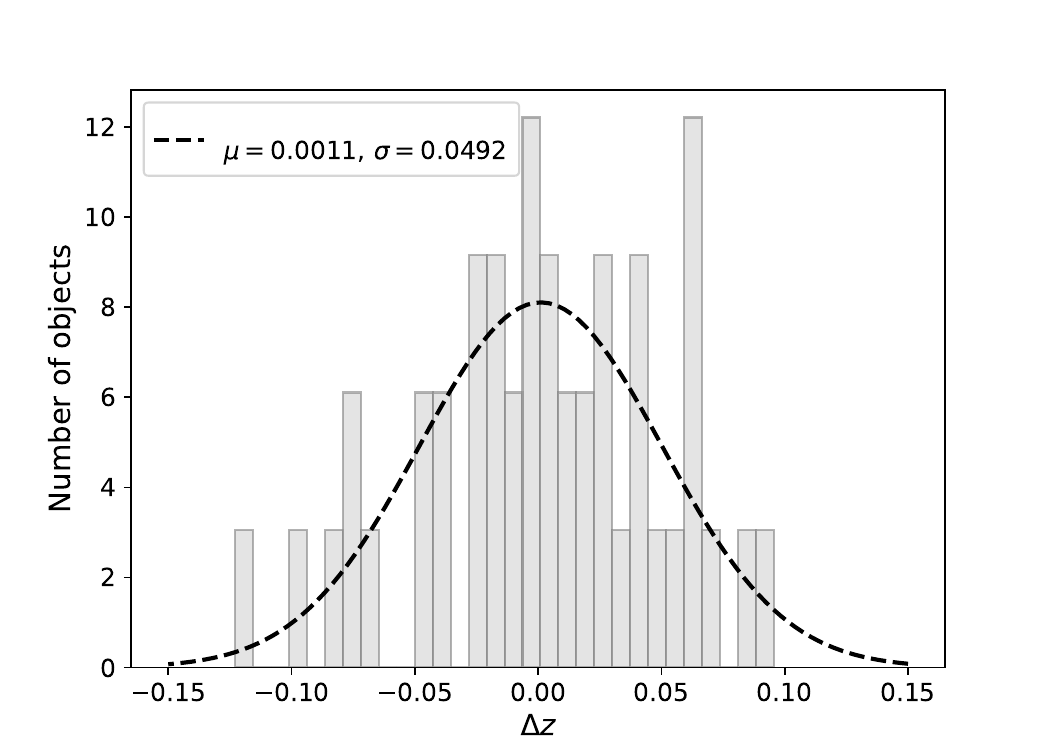}    
    \caption{Distribution of the redshift difference between each \INSPIRE\ UCMGs and the AMICO clusters at which they might be associated. The central value ($\mu$) and standard deviation ($\sigma$) of the Gaussian distribution fitted to the histogram are shown in the legend.}   \label{fig:deltaz_AMICO_histo}
\end{figure}

\subsection{Cluster membership from AMICO}
\label{sec:matchAMICO}
For each cluster candidate, AMICO  provides a list of galaxies with the probability of being a cluster member ($P_{\rm cluster}$). 
The $P_{\rm cluster}$ is computed through a cluster model, which assumes a luminosity function and a radial density profile. We note that the probability is distance-dependent in the sense that galaxies spatially more distant from the cluster centre will,  by construction, have a lower probability. 
We cross-match the list of galaxies that could be a member of one of the AMICO clusters with the 52 \INSPIRE\ UCMGs, 
finding that 45 out of the 52 \INSPIRE\ UCMGs have $P_{\rm cluster}>0$. 
However, only 9 of them have a probability of being members greater than or equal to 0.5 (i.e., $P_{\rm cluster}\ge0.5$), which means that their probability to be in a cluster is larger than their probability to be in the field. We will denote these as `safe' detections for the remainder of the paper.

In Fig.~\ref{fig:deltaz_AMICO_histo}, we show the distribution of the redshift difference, $\Delta z = (z_{\rm UCMG} - z_{\rm  cluster})/(1+z_{\rm UCMG})$, between each of the 45 UCMGs with a match and the corresponding AMICO cluster. The redshifts exhibit close proximity, with a mean $\Delta z = 0.0011$ and a standard deviation $\sigma = 0.0492$, as reported in the legend, although there is one object with $|\Delta z |>0.1$  (J0844+0148, $z_{\rm UCMG} = 0.2837$, $z_{\rm cluster} = 0.4413$). We stress that the clusters' redshifts are based on the photometric redshifts of their members. 

The results of the cross-matching between the AMICO cluster catalogue and \INSPIRE\ galaxies are listed in the second block of columns of Table~\ref{tab:environment_new}, where we list, for the 45 objects with a match, the redshift of the AMICO cluster compatible with that of the UCMG, the probability of the object to be a member of that cluster, the SNR of the detection, the level of the purity ($\mathscr{P}$), and the virial mass of the cluster ($M_{200}$). The seven objects that lack a match in AMICO are still listed in the table but without a numerical value for these quantities. In fact, for these, the cluster finding algorithm does not find a suitable association to any of the detected cluster candidates. This suggests that they might be field galaxies or reside in groups of galaxies that are too small to be detected by AMICO. 

\begin{figure}
    \centering  \includegraphics[width=1.08\columnwidth]{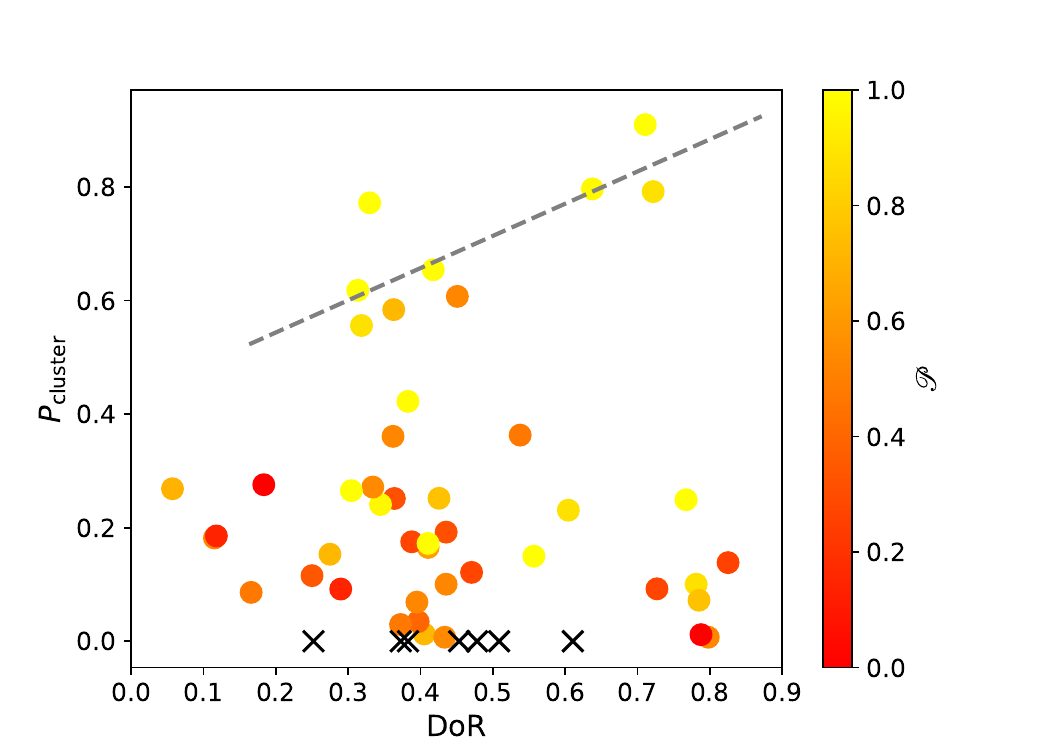}
\caption{Probability to belong to a cluster according to AMICO against the DoR for the 45 \INSPIRE\ UCMGs with a match in AMICO, colour-coded by the purity. The seven objects without a match are visualised as black crosses at $P_{\rm cluster}=0$. When restricting to `safe' detections (galaxies with $P_{\rm cluster} \ge 0.5$, see text for more details), a linear relation (grey dashed line) is found, with higher probabilities for objects with larger DoR.}
\label{fig:p_cluster}
\end{figure}

The 45 \INSPIRE\ galaxies matched with the AMICO cluster candidates exhibit a broad range in DoR, as visible from Fig.~\ref{fig:p_cluster}, where we colour-coded the data points by the purity of the cluster, $\mathscr{P}$. In the same figure, the remaining 7 UCMGs without a match are illustrated as black crosses at $P_{\rm cluster}=0$. A linear relation, showed as a grey dashed line, emerges when we restrict to `safe' detections, i.e., limiting to points with $P_{\rm cluster} \ge 0.5$, corresponding to a 50\% probability of belonging to the cluster identified by the algorithm. 

In summary, according to the analysis carried out from AMICO, 9 objects are members of a cluster of galaxies ($P_{\rm cluster}\ge0.5$), while 7 are most likely in the field with a high degree of confidence as they do not have a match. The remaining 36 objects exhibit a wide range of $P_{\rm cluster}$ values, but all lower than 0.5, hence preventing a safe classification of the local environment. Considering the grouping from \citetalias{Spiniello23}, all 9 extreme relics (DoR $>0.7$) have a match with an AMICO cluster. Of the 7 UCMGs lacking a match in AMICO, 6 are relics ($0.34 \le \rm{DoR} < 0.7$) and 1 is a non-relic (DoR$ <0.34$). There are 2 extreme relics, 4 relics, and 3 non-relics with $P_{\rm cluster}\ge0.5$. 
Henceforth, there appears to be no clear environmental preference for UCMGs with varying degree of relicness but that a linear correlation exists between DoR and $P_{\rm cluster}$ for the `safe' detections ($P_{\rm cluster}\ge0.5$). Furthermore, we notice that UCMGs with a very extended SFH (DoR $<0.3$) all show rather low probabilities of being in a cluster. 

\begin{table*}
\centering
\caption{Classification of the local environment for the 52 \INSPIRE\ UCMGs. Nine objects are in an over-dense region (top rows) with high degree of confidence, 17 definitively in under-dense environments (middle rows). For the remaining, only a tentative environment classification is provided. Within each environment group (horizontal lines), galaxies are ordered in descending order of DoR.
We list the \INSPIRE\ ID, the DoR and, the redshifts of the UCMGs ($z_{\rm UCMG}$) in the first vertical block (from \citetalias{Spiniello23}). In the second block, we list quantities derived from AMICO: redshift of the cluster ($z_{\rm cluster}$), the probability for the UCMG to belonging to that cluster ($P_{\rm cluster}$), the signal-to-noise ratio (SNR) of  detection, its purity ($\mathscr{P}$), the cluster's virial mass in $M_{\odot}$ ($M^{200}_{\rm cluster}$), and the logarithmic distance of the UCMG from the cluster centre in kpc ($\log D_{\rm A}$). Finally, in the third block, we list the over-density value ($\delta_{\Sigma}$), the logarithmic distance of each UCMG from its centre in kpc ($\log D_{\rm GZ}$), both derived from GaZNets, and lastly, the environment classification. Rows corresponding to objects without a match in AMICO or GaZNets are listed with a ---.}
\label{tab:environment_new} 
\begin{tabular}{ccc|cccccc|ccc}
\hline
\addlinespace[1.3mm] 
  \multicolumn{1}{c}{ID} &
   \multicolumn{1}{c}{DoR} &
   \multicolumn{1}{c|}{$z_{\rm UCMG}$} &
    \multicolumn{1}{c}{$z_{\rm cluster}$} &
  \multicolumn{1}{c}{$P_{\rm cluster}$} &
  \multicolumn{1}{c}{SNR} &
  \multicolumn{1}{c}{$\mathscr{P}$} &
  \multicolumn{1}{c}{$M^{200}_{\rm cluster}$} &
   \multicolumn{1}{c|}{$\log D_{\rm A}$} &
  \multicolumn{1}{c}{$\delta_{\Sigma}$} &
  \multicolumn{1}{c}{$\log D_{\rm GZ}$} &
  \multicolumn{1}{c}{Environment} \\
\multicolumn{1}{c}{INSPIRE} &
\multicolumn{1}{c}{} &
\multicolumn{1}{c|}{} &
\multicolumn{1}{c}{} &
\multicolumn{1}{c}{} &
\multicolumn{1}{c}{} &
\multicolumn{1}{c}{} &
\multicolumn{1}{c}{$[M_{\odot}]$} &
\multicolumn{1}{c|}{[kpc]} &
\multicolumn{1}{c}{} &
\multicolumn{1}{c}{[kpc]} &
\multicolumn{1}{c}{} \\
\hline
J0211-3155	&	0.72	&	0.3012	&	0.2714	&	0.79	&	4.84	&	0.88	&	13.505	&	2.52	&	5.36	&	2.78	&	C\\
J2359-3320	&	0.71	&	0.2888	&	0.2913	&	0.91	&	7.87	&	1.00	&	14.149	&	2.51	&	12.98	&	2.48	&	C\\
J0920+0212	&	0.64	&	0.2800	&	0.2913	&	0.80	&	5.85	&	0.98	&	13.871	&	2.66	&	11.59	&	2.86	&	C\\
J0314-3215	&	0.42	&	0.2874	&	0.2616	&	0.65	&	6.09	&	0.99	&	14.125	&	2.85	&	12.04	&	2.90	&	C\\
J0844+0148	&	0.45	&	0.2837	&	0.4413	&	0.61	&	3.96	&	0.53	&	13.528	&	2.44	&	4.64	&	3.10	&	C\\
J1202+0251	&	0.36	&	0.3298	&	0.3216	&	0.58	&	4.28	&	0.72	&	13.428	&	2.73	&	7.04	&	2.76	&	C\\
J1436+0007	&	0.33	&	0.2210	&	0.2517	&	0.77	&	6.30	&	1.00	&	13.647	&	2.58	&	3.32	&	3.16	&	C\\
J0904-0018	&	0.32	&	0.2989	&	0.3216	&	0.56	&	4.69	&	0.88	&	13.719	&	2.83	&	5.81	&	2.63	&	C\\
J1402+0117	&	0.31	&	0.2538	&	0.2616	&	0.62	&	7.31	&	1.00	&	14.089	&	3.05	&	17.06	&	3.00	&	C\\
 \hline							J1438-0127	&	0.78	&	0.2861	&	0.3013	&	0.10	&	4.48	&	0.88	&	13.266	&	3.40	&	2.51 &	3.06$^{+}$	&	F\\
J1412-0020	&	0.61	&	0.2783	&	---	&	---	&	---	&	---	&	---	&	---	&	2.10	&	2.71$^{+}$	&	F\\
J1457-0140	&	0.47	&	0.3371	&	0.2419	&	0.12	&	3.47	&	0.27	&	13.027	&	3.33	&	2.86	&	3.10$^{+}$ &	F\\
J2356-3332	&	0.44	&	0.3389	&	0.2813	&	0.10	&	3.12	&	0.53	&	12.701	&	3.29	&	0.54	&	3.14$^{+}$	&	F\\
J0918+0122	&	0.43	&	0.3731	&	0.4647	&	0.007	&	3.82	&	0.54	&	13.262	&	2.90	&	2.24	&	2.48$^{+}$	&	F\\
J1411+0233	&	0.41	&	0.3598	&	0.3427	&	0.17	&	4.06	&	0.62	&	13.724	&	3.21	&	1.51	&	2.46$^{+}$	&	F\\
J1420-0035	&	0.41	&	0.2482	&	0.3427	&	0.17	&	5.72	&	0.99	&	13.916	&	3.12	&	2.37	&	1.85$^{+}$	&	F\\
J1114+0039	&	0.40	&	0.3004	&	0.4289	&	0.01	&	4.25	&	0.72	&	13.461	&	3.34	&	2.99	&	3.09$^{+}$	&	F\\
J0316-2953	&	0.40	&	0.3596	&	0.4647	&	0.03	&	3.66	&	0.40	&	13.178	&	3.22	&	2.81	&	2.43$^{+}$	&	F\\
J1527-0012	&	0.38	&	0.4000	&	---	&	---	&	---	&	---	&	---	&	---	&	2.35	&	2.32$^{+}$	&	F\\
J1527-0023	&	0.37	&	0.3499	&	---	&	---	&	---	&	---	&	---	&	---	&	2.81	&	3.14$^{+}$	&	F\\
J0321-3213	&	0.37	&	0.2947	&	0.3537	&	0.03	&	3.70	&	0.47	&	13.516	&	3.47	&	1.45	&	2.48$^{+}$	&	F\\
J2257-3306	&	0.27	&	0.2575	&	0.2913	&	0.15	&	4.26	&	0.72	&	13.108	&	3.27	&	2.72	&	3.05$^{+}$	&	F\\
J0920+0126	&	0.25	&	0.3117	&	---	&	---	&	---	&	---	&	---	&	---	&	2.29	&	3.09$^{+}$	&	F\\
J1142+0012	&	0.18	&	0.1077	&	0.0951	&	0.28	&	2.88	&	0.00	&	12.744	&	3.06	&	2.43	&	3.10$^{+}$	&	F\\
J0226-3158	&	0.12	&	0.2355	&	0.2913	&	0.19	&	3.04	&	0.14	&	12.981	&	3.03	&	2.56	&	2.93$^{+}$	&	F\\
J2327-3312	&	0.06	&	0.4065	&	0.2913	&	0.27	&	4.21	&	0.70	&	13.293	&	3.28	&	2.34	&	2.40$^{+}$	&	F\\
 \hline																						
J2305-3436	&	0.80	&	0.2978	&	0.2616	&	0.007	&	3.38	&	0.53	&	13.077	&	3.47	&	6.06	&	2.26	&	C$^{*}$\\
J2204-3112	&	0.78	&	0.2581	&	0.3652	&	0.07	&	4.59	&	0.76	&	13.255	&	3.27	&	5.53	&	2.89	&	C$^{*}$\\
J1040+0056	&	0.77	&	0.2716	&	0.2223	&	0.25	&	6.48	&	1.00	&	13.736	&	2.74	&	8.14	&	2.81	&	C$^{*}$\\
J1449-0138	&	0.60	&	0.2655	&	0.3114	&	0.23	&	4.81	&	0.88	&	13.595	&	3.17	&	6.57	&	3.14	&	C$^{*}$\\
J0838+0052	&	0.54	&	0.2702	&	0.1929	&	0.36	&	3.83	&	0.47	&	13.549	&	2.80	&	6.19	&	2.56	&	C$^{*}$\\
J0240-3141	&	0.43	&	0.2789	&	0.2714	&	0.25	&	4.62	&	0.76	&	13.598	&	3.22	&	9.61	&	3.17	&	C$^{*}$\\
J1228-0153	&	0.39	&	0.2973	&	0.1733	&	0.07	&	3.49	&	0.53	&	13.042	&	3.32	&	10.54	&	3.17	&	C$^{*}$\\
J1447-0149	&	0.38	&	0.2074	&	0.2321	&	0.42	&	5.69	&	0.99	&	13.438	&	3.11	&	10.77	&	3.17	&	C$^{*}$\\
J2312-3438	&	0.36	&	0.3665	&	0.3652	&	0.25	&	3.31	&	0.31	&	13.172	&	2.64	&	9.45	&	2.78	&	C$^{*}$\\
J1414+0004	&	0.36	&	0.3030	&	0.2223	&	0.36	&	3.92	&	0.53	&	13.113	&	3.00	&	5.33	&	3.18	&	C$^{*}$\\
J1154-0016	&	0.11	&	0.3356	&	0.2517	&	0.18	&	6.44	&	0.53	&	13.899	&	3.53	&	5.44	&	3.17	&	C$^{*}$\\

J0847+0112	&	0.83	&	0.1764	&	0.2027	&	0.14	&	3.36	&	0.26	&	12.963	&	3.24	&	3.68	&	2.32	&	F$^{*}$\\
J0909+0147	&	0.79	&	0.2151	&	0.1538	&	0.01	&	2.89	&	0.00	&	12.853	&	2.91	&	4.19	&	3.17	&	F$^{*}$\\
J0842+0059	&	0.73	&	0.2959	&	0.3216	&	0.09	&	3.35	&	0.27	&	13.035	&	3.34	&	3.48	&	2.46	&	F$^{*}$\\
J0224-3143	&	0.56	&	0.3839	&	0.3114	&	0.15	&	4.88	&	1.00	&	13.445	&	3.43	&	3.38	&	3.16	&	F$^{*}$\\
J0317-2957	&	0.51	&	0.2611	&	---	&	---	&	---	&	---	&	---	&	---	&	3.87	&	2.45	&	F$^{*}$\\
J2202-3101	&	0.48	&	0.3185	&	---	&	---	&	---	&	---	&	---	&	---	&	4.38	&	2.91	&	F$^{*}$\\
J1218+0232	&	0.45	&	0.3080	&	---	&	---	&	---	&	---	&	---	&	---	&	4.64	&	3.18	&	F$^{*}$\\
J0917-0123	&	0.44	&	0.3602	&	0.3013	&	0.19	&	3.45	&	0.31	&	12.292	&	3.11	&	4.07	&	2.18	&	F$^{*}$\\
J0857-0108	&	0.39	&	0.2694	&	0.1929	&	0.18	&	3.43	&	0.27	&	13.163	&	2.92	&	4.56	&	2.40	&	F$^{*}$\\
J1128-0153	&	0.34	&	0.2217	&	0.1831	&	0.24	&	5.13	&	0.97	&	12.798	&	3.34	&	3.54	&	3.17	&	F$^{*}$\\
J1417+0106	&	0.33	&	0.1794	&	0.1831	&	0.27	&	3.78	&	0.54	&	12.610	&	3.08	&	3.50	&	3.03	&	F$^{*}$\\
J1156-0023	&	0.30	&	0.2552	&	0.2616	&	0.27	&	9.02	&	1.00	&	14.421	&	3.40	&	3.22	&	3.16	&	F$^{*}$\\
J0326-3303	&	0.25	&	0.2970	&	0.2714	&	0.11	&	4.06	&	0.34	&	13.105	&	3.39	&	3.60	&	2.91	&	F$^{*}$\\
J1026+0033	&	0.29	&	0.1743	&	0.1440	&	0.09	&	2.91	&	0.14	&	12.918	&	3.28	&	---	&	---	&	F$^{*}$\\
J1456+0020	&	0.17	&	0.2738	&	0.3216	&	0.09	&	3.85	&	0.47	&	13.409	&	3.29	&	3.15	&	2.25	&	F$^{*}$\\
\hline  
\end{tabular}
 \begin{flushright}
 $^*$Tentative environment classification (see the text for more details).\\
 $^{+}$The distances have been calculated from the nearest density peak, which is, however, not significant compared to the background level.\\
 \end{flushright}
\end{table*}
\subsection{Over-densities identification from the GaZNets catalogue}
\label{sec:ML}
In this section, we describe the cross-match between the \INSPIRE\ catalogue and the GaZNets galaxy catalogue described in section~\ref{sec:gaznets}. We select, for each of the 52 UCMGs, all and only galaxies (objects with $P_{\rm gal}\ge 0.9$ in the catalogue by Feng et al., submitted) having redshift compatible to that of the \INSPIRE\ objects and $m_r<22$. 
In particular, for the redshift, we compute $\Delta z = (z_{\rm UCMG} - z_{\rm phot})/(1+z_{\rm UCMG})$ and retrieve objects with $| \Delta z | \le 0.03$. To justify our choices in magnitude and redshift range, we use spectroscopic data from the Dark Energy Spectroscopic Instrument (DESI) Survey \citep{levi2019dark, DESI2023}. We cross-match the catalogue with the GaZNets one and check the precision of the ML photometric redshifts against the spectroscopic ones. For objects brighter than $m_r<22$, the uncertainty in redshift estimation is of the order of 0.03, while it increases for fainter objects. 
Furthermore, we also point the reader to fig.~3 in \citet{Li22}, where the accuracy of the GaZNets ML photometric redshifts is estimated on the training sample of 20\,000 galaxies used to test the performances of the ML models. 
We restrict our analysis to galaxies with $m_r<22$, corresponding to detecting galaxies with \Mstar+5 at $z=0.1$ and \Mstar+2 at $z=0.4$ for the redshifts of the UCMGs. For each galaxy in INSPIRE meeting this criterion and having $| \Delta z | \le 0.03$, we generate a density map within a 10 Mpc radius. This is accomplished using KDEpy\footnote{\url{https://kdepy.readthedocs.io}} library that performs a Kernel Density Estimation (KDE) on both 1D and 2D data. Specifically, we employ an Epanechnikov kernel \citep{Epanechnikov} with a bandwidth of 0.2 Mpc. 
We extract the peak of the density within a radius of 1.5 Mpc ($\sim r_{200}$ for a cluster with $M^{200}_{\rm cluster} \sim 10^{14}-10^{14.5} M_\odot$), to check for the presence of nearby over-densities (next nearest neighbour, $\Sigma_{\rm NNB}$). 
We then select 50 random regions all located from 5 to 9 Mpc from the UCMG and all with radius of 1.5 Mpc. We compute the mean and standard deviation of this distribution ($\Sigma_{\rm BKG}\pm \sigma_{\Sigma_{\rm BKG}}$), using it as a background estimation. Practically speaking, with this procedure, we are estimating the significance of an over-density in the proximity of each UCMG by the quantity 
\begin{equation}
\delta_{\Sigma} = \dfrac{\Sigma_{\rm NNB} - \Sigma_{\rm BKG}}{\sigma_{\Sigma_{\rm BKG}}}.
\end{equation}

To determine the threshold in $\delta_{\Sigma}$ for which we expect that an over-density is significant, we repeat the procedure, while randomising  the galaxy positions in the field. The threshold is then defined as the average value of $\delta_{\Sigma}$ calculated in these randomized fields, which is $\delta_{\Sigma, \rm{min}}=3$. For one galaxy, J1026+0033, it has not been possible to compute the value of $\delta_{\Sigma, \rm{min}}$. This object lies in proximity to a luminous star, which has been masked in the catalogues, causing a non-homogeneous mapping of the proximity of the UCMGs. 

\begin{figure}
    \centering  \includegraphics[width=1.08\columnwidth]{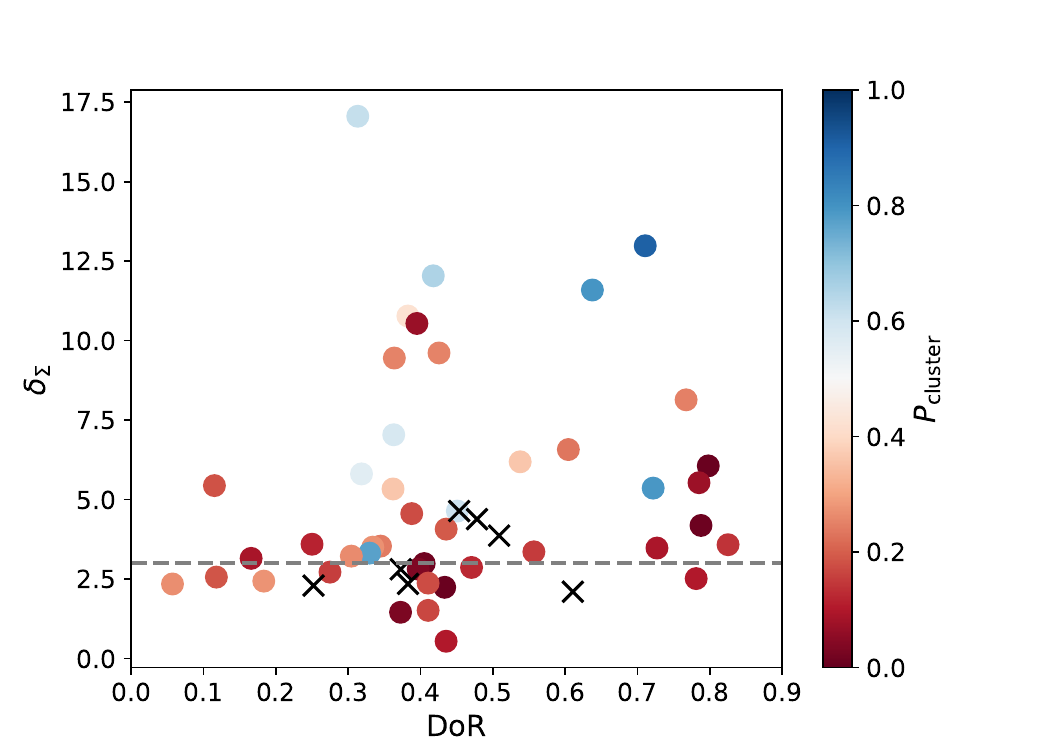}
\caption{Over-density as a function of the DoR for the 51 \INSPIRE\ UCMG for which a $\delta_{\Sigma, \rm{min}}$ could be computed. The dashed grey lines represent the threshold $\delta_{\Sigma, \rm{min}}$ for an over-density to be significant. The data points are colour-coded by their probability to reside in an AMICO cluster and the seven objects without a match in AMICO are visualised as black crosses.}
    \label{fig:Dor_nsigma_np}
\end{figure}

Figure~\ref{fig:Dor_nsigma_np} shows the over-density ($\delta_{\Sigma}$) values measured for each UCMG against their DoR. The dashed grey lines represent the $\delta_{\Sigma}$ threshold for an over-density to be significant. The data points are colour-coded by the $P_{\rm cluster}$ and those without a match in AMICO are visualised as black crosses.

We observe a similar behaviour to that described in the previous section (section~\ref{sec:AMICO}, Fig.~\ref{fig:p_cluster}). Nearly all UCMGs with extended SFHs (DoR $<0.3$) prefer under-dense environments. However, the scatter suddenly intensifies at intermediate DoR, where UCMGs reach values of $\delta_{\Sigma}$ as high as 17.06 and as low as $0.54$. 
The data points in the figure are colour-coded by the probability to reside in an AMICO cluster. All objects with $P_{\rm cluster}\ge 0.5$, i.e., `safe' detection, pass the $\delta_{\Sigma}$ threshold (grey dashed line). This results serves as a confirmation of the consistency between the two methods.

Interestingly, the vast majority of galaxies with DoR $\ge0.5$ are positioned above the $\delta_{\Sigma}$, suggesting, once again, that relics might slightly prefer over-dense environments, such as clusters of galaxies. Specifically, almost all (8 out of 9) extreme relics and two-thirds of relics with intermediate DoR values are above the threshold (with the maximum scatter), while roughly 50\% of the non-relics fall below the significance threshold. Only one out of the 14 non-relics has a very high value of $\delta_{\Sigma}$ (J1402+0117, $\delta_{\Sigma}=17.06$\footnote{From the map shown in Fig.~\ref{fig:map-news}, this galaxy seems to lie within two groups/clusters.}).  

Hence, according to GaZNets, 34 \INSPIRE\ UCMGs (65\% of the sample) reside in over-densities ($\delta_{\Sigma} \ge 3$). Of these, 8 (out of 9) are extreme relics, 8 are non-relics (out of 14), and only three (out of 9) have DoR $<0.3$.
 
\section{Results: characterisation of the local environment}
\label{sec:results}
In this section, we present the results of the analysis performed in the paper, focusing on the relation between the DoR and the environments occupied by the UCMGs. 
This topic was already introduced in previous papers (e.g., \citealt{Poggianti+13, Stringer+15_compacts, Damjanov+15_env_compacts, Tortora20}), which, however, lack any distinction between the ancient massive relics of the early Universe from relatively younger but equally compact and massive systems, that have undergone a much more extended SF. In the local Universe, \citetalias{Ferre-Mateu+17}, based on three confirmed relics, hinted for a correlation between the kinematics, structural, and stellar population parameters of three extreme relics and their local environment.  
The great step forward brought by \INSPIRE\ is the possibility to investigate on this matter with a much larger number statistics and covering a wider range in DoR. Hence, we have checked whether an environmental dependency exists for the stellar masses and sizes \citep{Scognamiglio20}, velocity dispersion values \citep{DAgo23}, and stellar parameters (IMF slope, metallicity, [Mg/Fe]; \citetalias{Maksymowicz-Maciata24}). Surprisingly, no statistically significant correlation was found, indicating that the larger \INSPIRE\ sample does not support the idea hinted at in \citetalias{Ferre-Mateu+17}. Velocity dispersion and stellar population parameters, although varying as a function of DoR, do not depend upon the density of the local environment in which a UCMG resides.

\begin{figure}
    \centering
    \includegraphics[width=0.95\columnwidth]{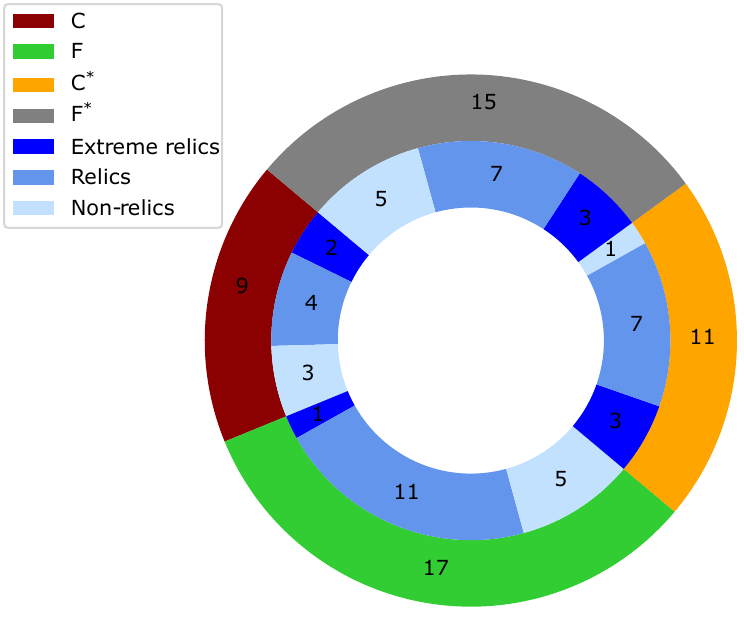}
    \caption{Summary of number of \INSPIRE\ UCMGs for each environment (outer circles), and DoR family (inner circles).}
    \label{fig:wedge}
\end{figure}

Table~\ref{tab:environment_new} provides a summary of the characterisation of the local environments for the 52 \INSPIRE\ UCMGs. 
The first (second) horizontal block of the table lists objects that are 
residing in an over-(under-)dense region with high degree of confidence, having very high (low) values of both $P_{\rm cluster}$ and $\delta_{\Sigma}$. 
Among the 26 UCMGs whose environment can be confidently determined (50\% of the sample), 9 are situated in clusters (labelled as `C') and 17 in the field (labelled as `F'). These galaxies exhibit a diverse range of DoR values in both environments. Notably, we confirm the absence of UCMGs with DoR $<0.3$ in cluster environments. 
Out of the remaining 26 \INSPIRE\ objects, 11 are likely to be cluster members, while 15 might inhabit either field environments or small galaxy groups, although a definitive determination remains uncertain. Specifically, the objects marked as `C$^{*}$' in Table~\ref{tab:environment_new} exhibit a significant $\delta_{\Sigma}$ value, exceeding the threshold by more than 1$\sigma$, yet they have an AMICO $P_{\rm cluster}$ value below 0.5. This implies that they reside within an over-dense region. Their low AMICO probability could stem from different factors, such as the UCMGs being situated relatively far from the cluster centre\footnote{We remind the reader that $P_{\rm cluster}$ is distance-dependent.}, or due to potential inaccuracies in the photometric redshift estimation of the cluster. Lastly, systems denoted with an `F$^{*}$' have a $\delta_{\Sigma}$ slightly above the threshold (but less than 1$\sigma$ away), along with a low $P_{\rm cluster}$ value, likely reside in the field or in the outskirts of a small galaxy group, with moderate density. 
The wedge plot in Fig.~\ref{fig:wedge} provides a graphical visualisation of the numbers of UCMGs based on the environment classification (outer circle) and on the three families defined in \citetalias{Spiniello23} according to the DoR (inner circle).

\begin{figure}
    \centering  \includegraphics[width=1.0\columnwidth]{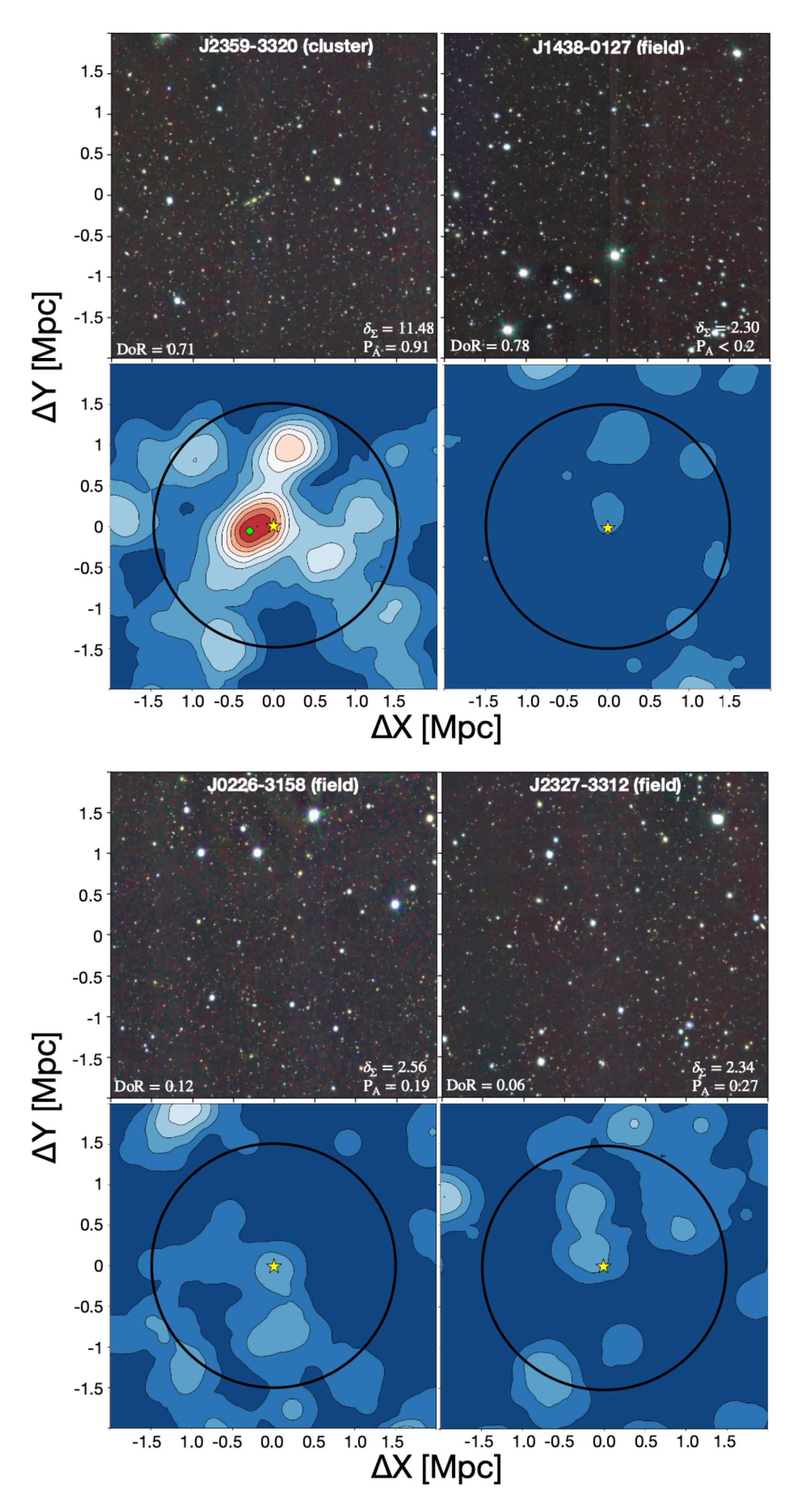}
    \caption{KiDS $gri$ cutouts (top panels) and KDE density maps (bottom panels) of size $2\times2$ Mpc$^{2}$ for the objects with highest (top row) and lowest (bottom row) DoR in the \INSPIRE\ sample, for which a definitive environment characterisation has been obtained. The cutouts also display DoR, $P_{\rm A}$, and $\delta_{\Sigma}$ values. The subscripts `A' is used to denote the $P_{\rm cluster}$ from `AMICO', for brevity and clarity in the images. In the KDE density maps, the yellow star indicates the position of the UCMG, while the green diamond identifies the cluster centre, if any. The circular aperture of 1.5 Mpc used to identify the nearest over-density is also drawn on the map in black.}
    \label{fig:map-extreme}
\end{figure}

\begin{figure*}
\includegraphics[width=\textwidth]{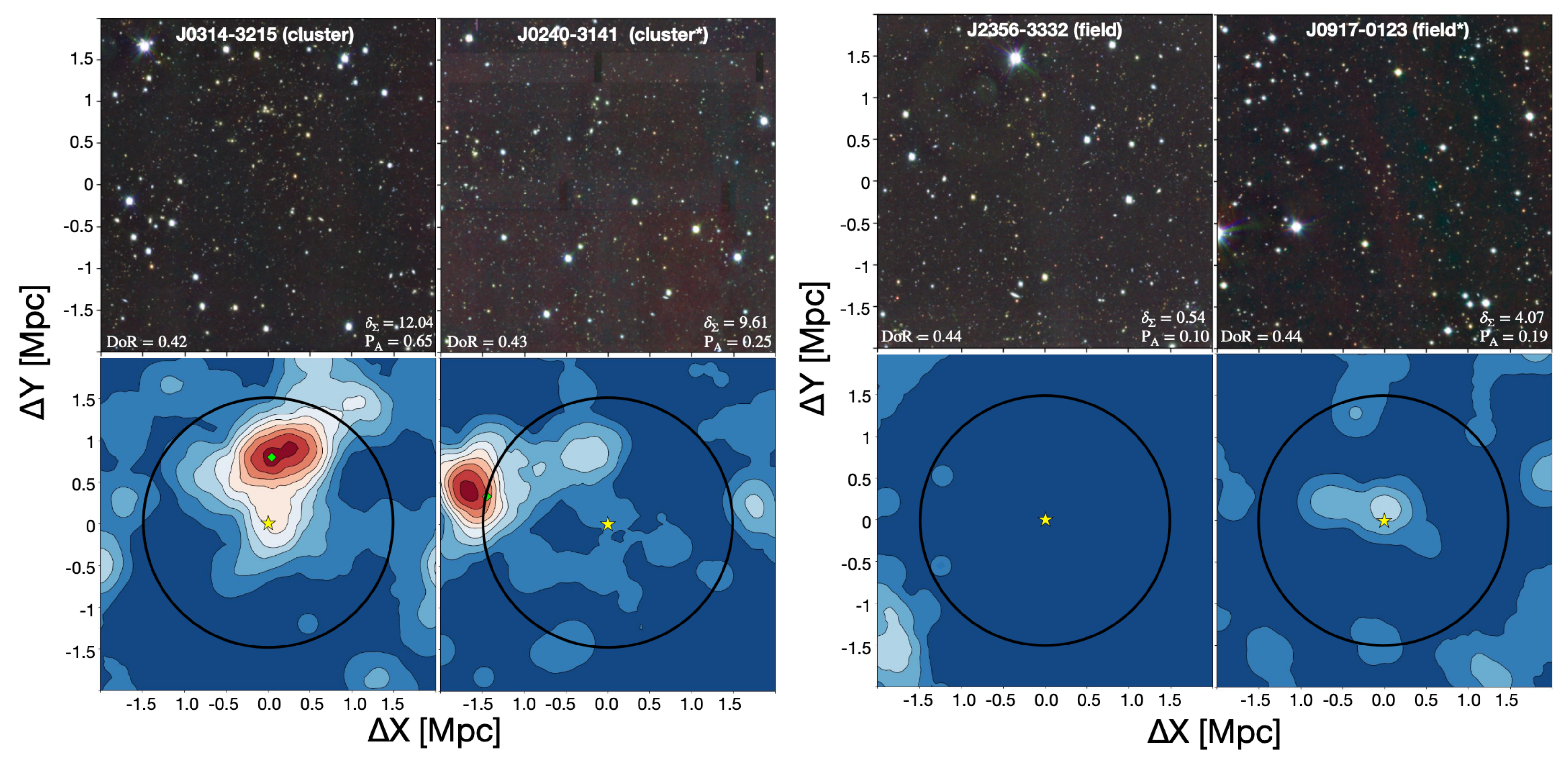}
    \caption{The same as in the previous figure, but for four \INSPIRE\ UCMGs with intermediate DoR residing in different environments (cluster on the left, field on the right). We display two `safe' classifications and two tentative ones. Symbols and scales are as in the previous figure.}
    \label{fig:map-interm}
\end{figure*}

\begin{figure*}
\includegraphics[width=\textwidth]{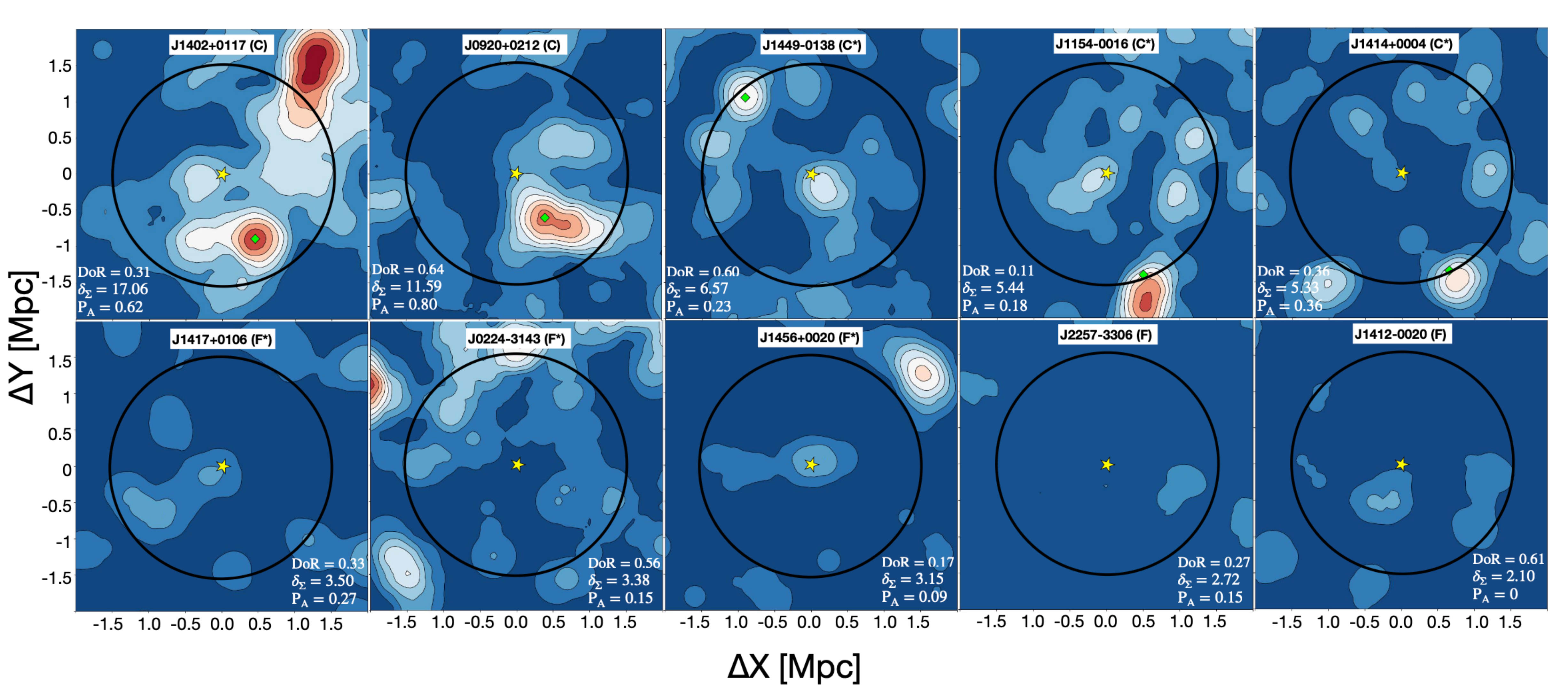}
    \caption{KDE density of $2\times2$ Mpc$^{2}$ for five objects in over-dense environments (top row) and five objects in under-dense environment (bottom row), ordered by their $\delta_{\Sigma}$ value. Symbols as in the previous figures.}
    \label{fig:map-news}
\end{figure*}

Figure~\ref{fig:map-extreme} shows 4 of the most extreme \INSPIRE\ objects (two highest and two lowest DoR), for which a definitive estimate of the local environment could be obtained (i.e., first two blocks of Table~\ref{tab:environment_new}). The two objects on the top row are two extreme relics with DoR $>0.7$, one of them are in a cluster one in the field. The bottom row depicts instead two objects with the lowest DoR, all in the field according to our classification. 
For each galaxy, the top panel illustrates the $gri$ colour-combined image from the KiDS survey while the bottom shows the KDE density map, both with size of $2\times2$ Mpc$^{2}$ and centred on the UCMG, which is plotted as a yellow star in the maps. For objects in clusters, we also highlight the position of the centre of the cluster candidate as a green diamond.  
Figure~\ref{fig:map-interm} shows instead four objects with very similar DoR (intermediate $\sim0.4$) but residing in different environments. The two top systems live in an over-density, while the two bottom ones inhabit under-dense environments. In all panels, the ID, DoR, $\delta_\Sigma$, and $P_{\rm cluster}$ (reported as $P_{\rm A}$, for brevity and clarity in the images) values are reported. Finally, Fig.~\ref{fig:map-news} shows the KDE maps for additional 10 objects, 5 in over-dense and 5 in under-dense environments to provide a general idea of the variety of the maps for the \INSPIRE\ sample. 
Objects in the same column (top and bottom) have very comparable DoR values. Moving from left to right within the same row, the $\delta_{\Sigma}$ values decrease.

\subsection{Correlation between the DoR and the UCMG distance to the cluster centre}
\begin{figure}
    \centering
\includegraphics[width=1.08\columnwidth]
{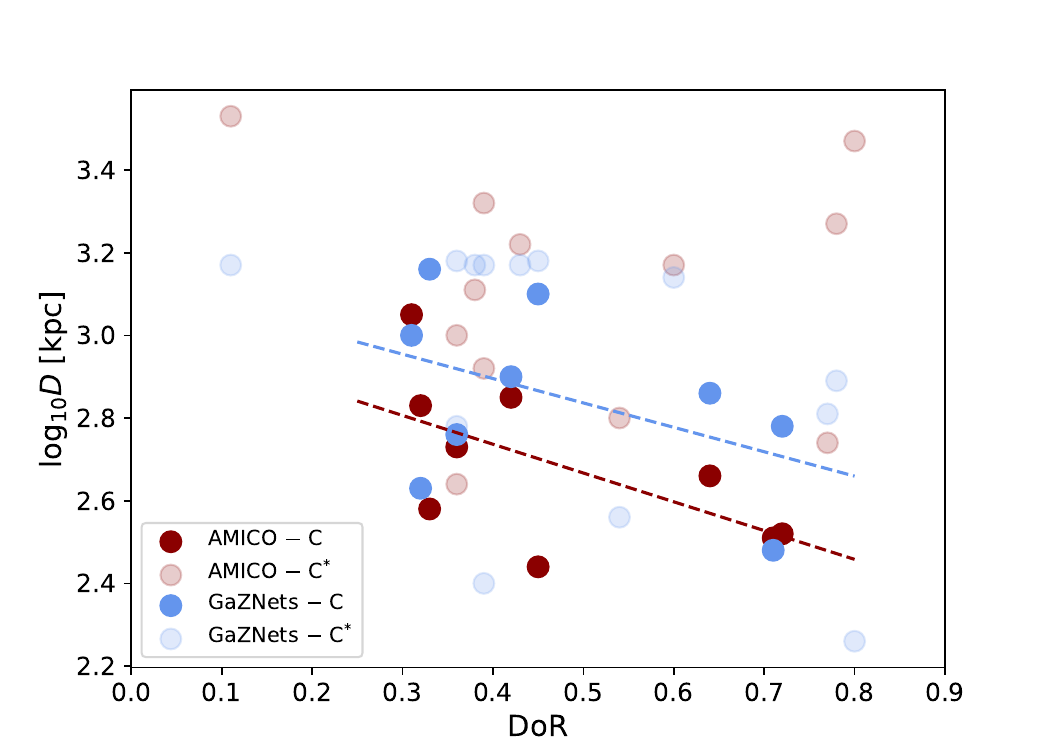}
\caption{Correlation between DoR and distances from a cluster candidate or an over-density, for AMICO (dark red points) and GaZNets (blue points), labelled with `C' in Table~\ref{tab:environment_new}. A linear fit is shown as dashed line for both cases. Lighter points show the UCMGs labelled as `C$^{*}$'  in Table~\ref{tab:environment_new}, which are not included in the fit as the classification of the environment is tentative.}
\label{fig:LogD_Dor}
\end{figure}

In Table~\ref{tab:environment_new}, we include the logarithmic values of the distance of an UCMG from both a cluster centre and from an over-density, in kpc. These values are visualized in Fig.~\ref{fig:LogD_Dor}, where they are plotted against the DoR. In this figure, the dark red points represent AMICO data, while the blue points denote GaZNets data. Galaxies confidently identified as belonging to a cluster or an over-density are labelled with `C', whereas those considered tentative members are marked with `C$^{*}$' (shaded circles). 
For both methods, a similar correlation is found: the higher the DoR, the closer to the cluster centre/over-density the UCMGs is located. Indeed, performing a linear fit between the aforementioned distances and the DoR values, a comparably negative correlation is evident in both cases: 

\begin{equation}
\begin{split}
\label{eq:amico}
\log_{10} D_{\rm AMICO} = 3.02 -(0.7\times \rm DoR) \\
\log_{10} D_{\rm GaZNets} = 3.13 -(0.6\times \rm DoR)   
\end{split}
\end{equation}
To assess the significance of the correlation between the DoR and the log$_{10}D$ for the `safe' detections, we perform a bootstrap analysis. We randomize the DoR and log$_{10}D$ values independently for 49 iterations and include the original data as the 50th iteration. The mean slope from these bootstrapped datasets is $\mu=0.01 \pm 0.46$ for AMICO and $\mu=-0.04 \pm 0.43$ for GaZNets. In contrast, the slopes from the original data are 0.7 for AMICO and 0.6 for GaZNets. These significant differences indicate that the observed correlations are not due to random fluctuations, affirming the robustness and significance of the relationship between DoR and log$_{10}D$.

It is worth noting that the physical significance of the distance from the centre varies significantly depending on whether one considers a very massive cluster of galaxies, which can extend up to much larger distances or small groups, that are instead generally much smaller in size. Nevertheless, from Table~\ref{tab:environment_new} it is clear that the cluster candidates identified by AMICO do not exhibit a wide range in total mass,  
spanning from 
$4.1\times10^{12}M_{\odot}$ to $2.6\times10^{14}M_{\odot}$, and with a mean and standard deviation of $(2.6\pm 0.1)\times10^{13}M_{\odot}$.
Finally, we also note that the distance-DoR relation shows a much larger scatter and a shallower slope when considering all the UCMGs that are tentatively residing in a cluster (i.e., C$^{*}$, shaded dark red points in Fig.~\ref{fig:LogD_Dor}). However, as already stressed, the quantities derived by AMICO are based on the detection algorithm that depends on less-precise photometric redshifts and on the analytical cluster model based on a luminosity function and a radial density profile. Furthermore, sometimes a system could be associated (through the membership probability) to more than one cluster. Hence, the estimate of the $\log D$ for AMICO, especially for tentative classification, could be biased or influenced by unaccounted factors. The over-density values from GaZNets are based on more accurate ML redshifts and simply trace the number of galaxies in the proximity of each UCMG, without assuming any model or light distribution.
\section{DISCUSSION} 
\label{sec:discussion}
From this study it has emerged that UCMGs are present in both over- and under-dense regions, in line with what previously reported \citep{Damjanov+15_env_compacts, Tortora20}. We further found that there is no correlation between their DoR and the density of the local environment. Our observational findings are in perfect agreement by results from simulations \citep{Moura2024}, which have shown that both relics and younger UCMGs exhibit no distinct preference towards either high- or low-density environments at $z=0$. Instead, they are distributed across a range of densities, as evidenced by their presence in various environments. 

In this section we focus on possible evolutionary scenarios that are able to explain the presence of UCMGs with all DoR both in over- and under-dense regions and possibly, the lack of objects with DoR$<0.3$ in cluster environment. 
We start by focusing on the most extreme relics, hypothesising on how they could avoid any interaction with other cluster members or very quickly stop forming stars in the field. We then focus on UCMGs with a very extended SFH, which seem to prefer under-dense environments. Finally, we stress that a possible third scenario might exist for UCMGs with low and intermediate DoR. Indeed, the observed near-by UCMGs might also be galaxies that had a three phase formation and evolution scenario. After the size growth phase they might have gone through a subsequent phase of re-compaction due to stripping and gas removal \citep{Kapferer2009, Peluso_2022,goller2023}, and are therefore observed to be ultra-compact today. This is highly unlikely for extreme relics, where the entire totality of the stellar population is almost as old as the Universe but it cannot be excluded for the remaining cases. 


\subsection{Extreme relics in clusters and in the field}
Among the 9 extreme relics ($\rm{DoR}> 0.7$) which have formed all their stellar mass at high-$z$, two are in an over-dense environment and one is in an under-dense region (J1438-0127, DoR $=0.78$) with a high degree of confidence. For the remaining 6 systems, 3 are likely to reside in clusters or groups and 3 more likely to be in the field.  
Moreover, when extreme relics are in a cluster, they tend to be closer to its centre than UCMGs with lower DoR. 

While it is quite intuitive to understand how a high-$z$ isolated, field red nugget can evolve passively and undisturbed through cosmic time without interacting with any other system, it is difficult to explain how it can survive untouched for many Gyr in the central regions of a cluster/group of galaxies.  
A plausible  explanation is that extreme relics have a very deep potential well and incredibly high density, which make them less susceptible to interactions with other galaxies (see also \citealt{Poggianti+13}). This scenario is also supported by the fact that extreme relics have a larger stellar velocity dispersion (up to $\sim 400$ \kms\ at $M_{\star}\sim10^{11} M_{\odot}$) than non-relics, both younger UCMGs and normal-sized ETGs, of similar stellar mass \citep{Ferre-Mateu+17,Spiniello+21,Spiniello23, Grebol23}. As found in \citet{PeraltadeArriba+16}, the high-velocity dispersion values and the hot ICM can prevent the growth of an accreted stellar envelope through mergers.  Furthermore, a dense environment could help in preventing a continuous SF. Indeed, stripping of the cold gas surrounding a galaxy might happen by the hot, dense diffuse intra-cluster gas, causing SF to cease \citep{Ma_2008, Bekki_2013, Roediger_2014, Steinhauser_2016, Foltz_2018}.  Finally, the so-called “galaxy harassment”, which is the combined effect of gravitational interactions between galaxies \citep{Merritt1983, Bialas2015} and their interaction with the potential well of the cluster as a whole \citep{Byrd1990}, has often been proposed to explain the formation of red sequence galaxies (e.g.~\citealt{Boselli_2014}). So, if the UCMGs entered in the cluster very early-on in time, or formed with it, they stopped forming stars and then evolved passively thereafter.

Our findings align well with the theoretical scenario described in \citet{Moura2024}. The authors have shown that, although in the local Universe relics and younger compact galaxies can be found in all environments, at $z>2$ the number of relics in clusters is higher when compared to younger UCMGs at the
same mass range (see also \citealt{Kimmig23}). Hence red nuggets that will be  
progenitors of relics were preferentially in an over-density at high-$z$. On the contrary, ultra-compact non-relics in a cluster entered in this dense environment only at later stages. 


In low-density environments, instead, red nuggets are less likely to merge and increase the `ex-situ' fraction of stars. Therefore we expect to find these ultra-compact systems still as such in more isolated environment \citep{PeraltadeArriba+16, Kimmig23,Moura2024}. However, if the red nuggets are surrounded by a gas reservoir, they might keep forming stars, although at a very low-rate until they consume it. Hence, extreme relics in the field might be the local descendants of red nuggets have been formed in gas-poor under-dense regions of the Universe at early epochs. 
This might be the case of J1438-0127, the only \INSPIRE\ extreme relic (DoR $=0.78$) in the field, and Mrk1216, the only near-by extreme relic definitively the field.

\subsection{Why do UCMGs with more extended SFHs prefer the field?}
At the other extreme of the distribution, we have found no objects with DoR $\le0.3$ definitively located in an over-dense region. Looking at the tentative classifications (marked with an asterisk (*) in Table~\ref{tab:environment_new}), only one out of ten UCMGs with such lower DoR could be in a cluster. Among the remaining objects, 6 are definitively in under-dense regions and 3 are tentatively in the field. 
These objects, although passive and relatively old ($\sim4-6$ Gyr) in an integrated sense, are characterised by a much more extended SFH and they are still forming a small percentage of their stellar mass today or stopped very recently, similar to those originally found in \citet{Trujillo+09_superdense} and \citet{Ferre-Mateu+12}. 
The preference for UCMGs with extended SFHs to live in the field could be explained by the fact that in these under-dense environment, major (and especially dry) mergers are rare (e.g., \citealt{de_Ravel_2009, Fakhouri_Ma_2009, Darg2010, Lin_2010, Kampczyk_2013, Ellison2013}). Major mergers can indeed trigger intense bursts of SF, thus shortening the overall SFH of galaxies \citep{Mihos1996, Gabor_2010, Man_2018}. In contrast, these red nuggets may have had a large reservoir of surrounding gas, which is gradually turned into stars over a much longer timescale.
If this scenario is true, then one should expect to find an anti-correlation between the H\textsc{I} column density and the DoR, which is however still too challenging to measure with the current radio telescopes. 


\section{Summary and Conclusions} \label{sec:conclusion}
In this seventh paper of the \INSPIRE\ survey, we have investigated whether a correlation exists between the DoR and the local environment for the 52 UCMGs in the sample, as originally hinted at in \citetalias{Ferre-Mateu+17}. 
We have started with cross-matching the \INSPIRE\ catalogue with a catalogue of cluster candidates from the KiDS survey, obtained using the AMICO algorithm \citep{Bellagamba18, Bellagamba19, Maturi19}. Out of the 52 UCMGs, 45 are associated with a cluster as potential members, with a wide range of probabilities, SNR of the cluster detection, and cluster purity values.

We have estimated the local density and its significance with respect to the background around each galaxy in \INSPIRE.
To this aim, we have utilised machine-learning-based photometric redshift from the GaZNets-1 catalogue \citep{Li22} and classification into quasars, galaxies, and stars from Feng et. al (submitted). This allowed us to identify galaxies with $m_r<22$ situated within 1.5 Mpc of each \INSPIRE\ UCMG, with a redshift difference of $\Delta z < 0.03$. 
We have then created a density map for each UCMG and estimated its significance over the background local density, obtained selecting 50 random region located around the UCMGs but at larger distances. 
We have finally assigned, on the basis of the analysis described above, two most likely environments (cluster or field) to each system in the \INSPIRE\ sample. 

These are the main results from the analysis presented in this paper: 
\begin{itemize}
\item UCMGs can be found in all kind of environments, from dense clusters of galaxies to under-dense field environments.
Taking the classifications in Table~\ref{tab:environment_new} at face value, we have found 20 UCMGs in a cluster and 32 in the field. Of these systems, 9 are in a cluster and 18 are in the field with high confidence environment classifications, while the remaining are only tentative\footnote{\citet{Tortora20} have found that the fraction of UCMGs in the field is slightly higher compared to that in clusters (see the right panel of their fig.~1).
We confirm this result and remind the reader that the \INSPIRE\ sample has been originally drawn from the same sample used in \citet{Tortora20}.}. 
\item There is no clear correlation between the DoR and the probability for a UCMG to reside in a cluster for the entire sample. However, when restricting the analysis to the `safe' detections (i.e., $P_{\rm cluster}\ge0.5$) a linear relation emerges: the most extreme relics exhibit the highest $P_{\rm cluster}$ values, as illustrated in Fig.~\ref{fig:p_cluster}.  
\item The density of the local environment where the UCMGs reside does not correlate with the DoR. However, UCMGs with an extended SFH (DoR $<0.3$) tend to prefer less dense environments, as shown in Fig.~\ref{fig:Dor_nsigma_np}. In fact, none of the objects with DoR $\le0.3$ reside in a dense environment with high confidence. Only one could tentatively be in a group/cluster (J1154-0016).  
\item A correlation is found between the DoR and the distance from the cluster centre log$_{10}D$, for both techniques. Indeed, Fig.~\ref{fig:LogD_Dor} illustrates that the higher the DoR, the closer the systems are
to the centre of the over-density. Additionally, the distances are comparable for the two methods ($2.6 <\log_{10} D<3.2$ kpc), at least for the high confidence classifications. 
\end{itemize}

In conclusion, our findings suggest that while a weak dependency on DoR exists, relics can be found across diverse environments with different local densities, consistent with both previous observational studies (\citetalias{Ferre-Mateu+17}; \citealt{Siudek2023}) and hydro-dynamical simulations \citep{PeraltadeArriba+16, Flores-Freitas22, Moura2024}. 
However, younger UCMGs, with a $\rm{DoR} < 0.3$ and thus characterised by a more extended SFH, are preferentially found in under-dense environments and reside almost exclusively in the field. 
We argue that these results are justified by the premise that if a red nugget has formed in an over-density at high redshift (or moved in quite early-on in cosmic time), the hot and dense ICM within the cluster gravitational potential has exerted high pressure and removed all the gas, hence quenching its SF activity \citep{PeraltadeArriba+16, Boselli_2022}. 
In clusters, extreme relics did not interact with other members, because they possess very deep potential wells, also indicated by their large stellar velocity dispersion values, and incredibly high densities. 

Conversely, in under-dense environments, red nuggets at high-$z$ that do not interact via mergers (growing in size) either keep forming stars at very low-rate consuming their gas envelope and becoming non-relics UCMGs, or they do not have any gas surrounding them and hence evolve passively and undisturbed without companions to merge with, nor gas reservoir to form new stars.  

Finally, we highlight that a fraction of the non-relics \INSPIRE\ UCMGs in a cluster might also have been originated by stripping phenomena on an originally larger massive system, that have caused a compaction phase at later cosmic epochs \citep{Dekel14, vanDokkum_2015}.

Only by further extending the number of fully characterised and spectroscopically confirmed UCMGs, studying their stellar populations in great detail, we will be able to quantify the number densities of the different types of ultra-compact objects and study their evolution with redshifts. Current and up-coming surveys like \textit{Euclid} \citep{Laureijs11}, \textit{Vera C. Rubin} Observatory's Legacy Survey of Space and Time (\textit{Rubin}-LSST; \citealt{2019ApJ...873..111I} , Multi-Object Spectroscopic Telescope (4MOST; \citealt{deJong_4MOST}), and WEAVE at William Herschel Telescope \citep{Jin_2023} will be transformational in extending the number statistics and pushing the redshift boundaries between the near-by and the high-$z$ Universe. 

\section*{Data Availability}
The \INSPIRE\ data used in this paper are publicly available via the ESO Phase 3 
Archive Science Portal under the collection \INSPIRE\ (\url{https://archive.eso.org/scienceportal/home?data_collection=INSPIRE}, \url{https:https://doi.eso.org/10.18727/archive/36}).

\section*{Acknowledgements}
The research was carried out at the Jet Propulsion Laboratory, California Institute of Technology, under a contract with the National Aeronautics and Space Administration (80NM0018D0004), © 2024. All rights reserved. DS is supported by JPL, which is operated under a contract by Caltech for NASA.  CS and CT acknowledge funding from the INAF PRIN-INAF 2020 program 1.05.01.85.11. GD acknowledges support by UKRI-STFC grants: ST/T003081/1 and ST/X001857/1. 



\bibliographystyle{mnras}
\bibliography{biblio_INSPIRE} 

\bsp	
\label{lastpage}
\end{document}